# 3D-Printed Surface Architecture Enhancing Superhydrophobicity and Viscous Droplet Repellency

*Gustav Graeber[1], Oskar B. Martin Kieliger[1], Thomas M. Schutzius[1,*], Dimos Poulikakos[1,*]*

[1] Laboratory of Thermodynamics in Emerging Technologies, Department of Mechanical and Process Engineering, ETH Zurich, Sonneggstrasse 3, CH-8092 Zurich, Switzerland.

[*] To whom correspondence should be addressed.

Prof. Dimos Poulikakos
ETH Zurich
Laboratory of Thermodynamics in Emerging Technologies
Sonneggstrasse 3, ML J 36
CH-8092 Zürich
SWITZERLAND
Phone: +41 44 632 27 38
Fax: +41 44 632 11 76
dpoulikakos@ethz.ch

Dr. Thomas M. Schutzius
ETH Zurich
Laboratory of Thermodynamics in Emerging Technologies
Sonneggstrasse 3, ML J 38
CH-8092 Zürich
SWITZERLAND
Phone: +41 44 632 46 04
thomschu@ethz.ch

**Abstract**

Macro-textured superhydrophobic surfaces can reduce droplet-substrate contact times of impacting water droplets; however, surface designs with similar performance for significantly more viscous liquids are missing, despite their importance in nature and technology such as for chemical shielding, food staining repellency, and supercooled (viscous) water droplet removal in anti-icing applications. Here, we introduce a deterministic, controllable and up-scalable method to fabricate superhydrophobic surfaces with a 3D-printed architecture, combining arrays of alternating surface protrusions and indentations. We show a more than threefold contact time reduction of impacting viscous droplets up to a fluid viscosity of $3.7 \text{ mPa} \cdot \text{s}$, which equals 3.7 times the viscosity of water at room temperature, covering the viscosity of many chemicals and supercooled water. Based on the combined consideration of the fluid flow within and the simultaneous droplet dynamics above the texture, we recommend future pathways to rationally architecture such surfaces, all realizable with the methodology presented here.

# Introduction

Great strides are being made toward developing superhydrophobic and icephobic surfaces, as their ability to stay dry, clean or ice-free is desired in a multitude of applications.[1–3] Considering fluid dynamical aspects, a common approach to study the performance of superhydrophobic surfaces is to carry out droplet impact experiments with millimeter-sized droplets, where the contact time and the impalement resistance are the two major figures of merit.[4,5] One can distinguish between macro-textured superhydrophobic surfaces that have surface features larger than about 0.1 mm, and macroscopically smooth surfaces that possess exclusively smaller roughness features in the micrometer- and nanometer-scale. Macroscopically smooth, micro-/nanotextured, superhydrophobic surfaces have shown great performance in resisting impalement,[6,7] while the lower limit of the droplet contact time on such macroscopically smooth surfaces is given by about 2.6 times the inertial-capillary time scale of the droplet.[8] A successful pathway to further reduce the contact time and undercut this lower limit of macroscopically smooth surfaces is the introduction of surface architectures with rationally designed macroscopic features of the size of about 0.1 mm. Impressive contact time reductions are accomplished by either achieving droplet rebound before recoiling and while at a stretched pancake shape[9–11] or by asymmetric rebound, or splitting the impacting droplet during rebound with the help of dedicated macrostructures.[12–14]

Here we fabricate superhydrophobic surfaces in a precise, facile and up-scalable manner by combining 3D mold printing, soft lithography with flexible polymeric materials and spray coating. The surfaces possess a hierarchical architecture composed of a macro-texture decorated with micro-nanoscale roughness. The proposed textures consist of two distinct and complementary macroscopic features, namely, protruding and indented truncated cones alternating in a regular square array. This specific design enables high mechanical robustness due to the short cone height, while still maintaining a high surface area, which is required for droplet repellency. Our fabrication approach allows for a large-area reproduction in future work using a step and repeat imprint process. The flexibility of the employed material enables that the final surface can be imparted on a variety of substrates including curved surfaces. Our surface fabrication approach presented herein also paves the way for further testing a variety of textured surfaces with different complex texture geometries for a multitude of applications.

We experimentally probe the performance of our surfaces in a droplet property regime beyond the popular pure water at atmospheric conditions. We explore the surface behavior for viscous fluids, which is relevant for chemical shielding, where prolonged contact time is associated with detrimental outcomes. Also, as the viscosity of water increases drastically when cooling it,[15] studying viscous model fluids is important in the development of icephobic surfaces, which can repel supercooled water droplets before the water can freeze and stick to the surface.[2,7]

**Materials and Methods**

We fabricated the mold for the macro-textured surface employing a high precision 3D printer Nanoscribe system, which uses two-photon polymerization of photo-curable resists to form three-dimensional structures with near nanometer-level precision. To reduce the time needed to print the structure, we followed an optimized printing process whereby only the surface and a mechanically stable scaffold were polymerized, while liquid resist in the interior of the structure within the scaffold was left not polymerized. After printing, the structure was developed in 1-methoxy-2-propanol acetate (PGMEA) and exposed to ultraviolet radiation to cure the remaining liquid resist trapped within the cured surface of the scaffold. Subsequently, the printed mold was exposed for 20 s to a $C_4F_8$ plasma to facilitate the mold removal during the following soft lithography step. Figure 1(a) and (b) show a schematic of the finished mold. It has a total diameter of 8.5 mm with a surface architecture consisting of a regular square array of alternating protruding truncated cones and inversed identically shaped indentations. The cone height, base diameter, and top diameter are $300\,\mu\text{m}$, $240\,\mu\text{m}$, and $120\,\mu\text{m}$, respectively. In Figure 1(b) we also introduce the opening radius $r_{open} = 120\,\mu m$ of the indentations, which serves as a reperesentative length scale in the subsequent fluid flow analysis. The printed mold was characterized in a scanning electron microscope (SEM). Figures 1(c) and (d) show SEM micrographs of the printed mold and an unfinished surface with shell and scaffold, respectively. Figure 1(e) shows a micrograph of polydimethylsiloxane (PDMS, mixing ratio 10:1 monomer to curing agent) that was cast in the mold and removed demonstrating the ease of replication. For control experiments we used macroscopically smooth microscope glass slides (VWR, cut edge, ECN 631-1550) and flat, untextured PDMS. To render the studied surfaces superhydrophobic, we first created a dispersion consisting of poly(tetrafluoroethylene) (PTFE; Sigma-Aldrich, powder, 1 µm particle size) and acetone. The concentration was 3 wt.% PTFE in acetone. We probe-sonicated (Sonics Vibracell, VCX-130, 130 W, 60 % amplitude, 20 kHz frequency, 3 mm probe diameter) the dispersion three times for 10 s to disperse the PTFE micro-particles, and deposited the dispersion onto the surfaces with a siphon-feed airbrush (Paasche, VL 0316, 0.73 mm head, air back-pressure 3 bar). After spray coating, the surfaces were heated to $130\,°C$ for two hours to remove the solvent and to promote adhesion between the coating and the substrate. Figures 1(f) and (g) show the final macro-textured surface and a magnified image of the coating, respectively. We refer to both the macroscopically smooth PTFE-coated glass slide sample and the macroscopically smooth PTFE-coated, untextured PDMS as "S-PTFE" (with S for smooth), because they showed no discernable differences in performance. We term the macro-textured, PTFE-coated surface with protrusions and indentations as "T-PTFE" (with T for textured). The surface identifiers are summarized in Table 1.

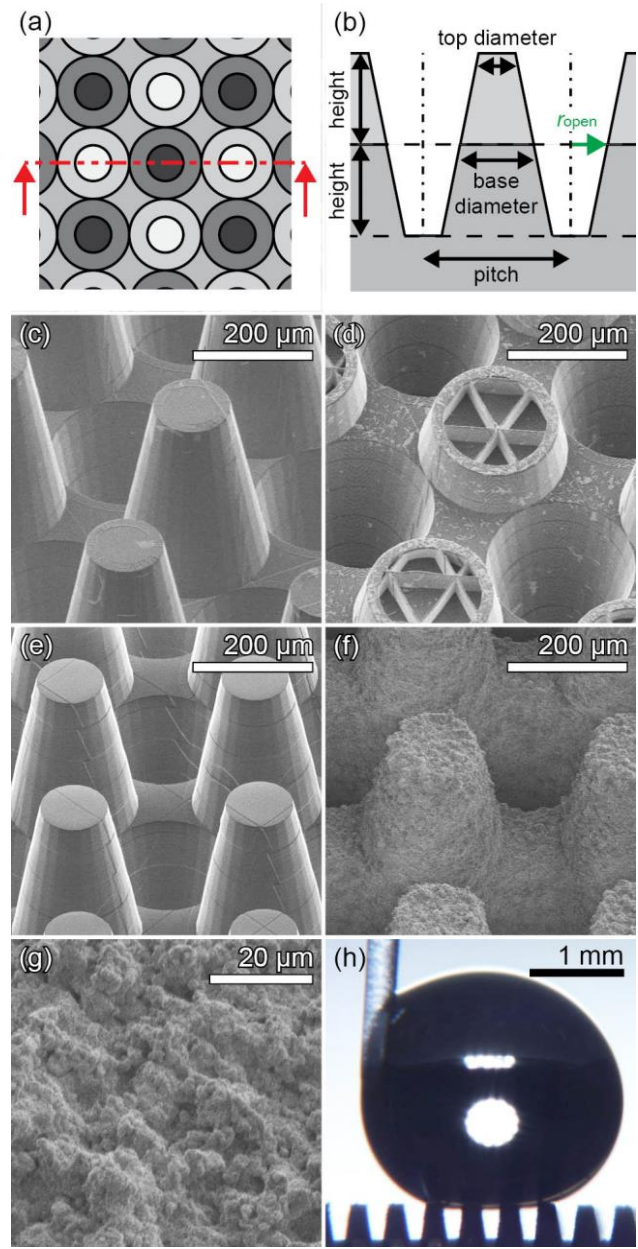

**Figure 1. Properties and fabrication of the macro-textured surface.** (a) Top-view sketch showing the protruding (bright) and indenting (dark) surface features. (b) Section-view along the red line in (a) introducing the texture opening radius $r_{\text{open}}$. (c) SEM micrograph of the 3D printed mold used to generate the macro-texture. (d) SEM micrograph showing an intermediate step during 3D printing of the mold with shell and scaffold. (e) PDMS reproduction of the mold. (f) PTFE spray coating applied to the PDMS reproduction. (g) Magnified view of the PTFE spray coating. (h) Side-view of a water droplet on the T-PTFE surface during contact angle measurement.

We prepared another coating system for reference (control) experiments.[16] Separate solutions of 10 wt.% poly(vinylidene fluoride) (PVDF) in 1-Methyl-2-pyrrolidinone (NMP) and 10 wt.% poly(methyl methacrylate) (PMMA) in acetone were prepared by dissolving the polymers under mechanical mixing for 12 hours at 50 °C and at room temperature, respectively. The dispersion consisted of 2.3 g of hydrophobic fumed silica (HFS, Aerosil R 8200, Evonik, powder, $\approx 10$ nm particle size), which was mixed with 33.8 g acetone using probe sonication three times for 30 s. 2.0 g of the previously prepared PVDF solution and 2.0 g of the PMMA solution were added to the HFS acetone suspension and were mechanically shaken at room temperature yielding the final dispersion. This final dispersion was applied by airbrush onto both the macro-smooth and the macro-textured substrates to render them superhydrophobic. We call the resulting samples "S-HFS" and "T-HFS" (see Figure S1 for SEM micrographs of the T-HFS surface). We used two different coating systems, because the PTFE coating is able to repel a broad range of liquids including glycerol, while the HFS coating is optimized for best water impalement resistance during droplet impact but cannot repel glycerol. Figure 1(h) shows a water droplet resting on the T-PTFE surface during contact angle measurements. The surface identifiers, the coating system and the water advancing and receding contact angles of the prepared surfaces are summarized in Table 1.

**Table 1. Properties of the prepared surfaces.** In the surface identifiers, "S" means macroscopically smooth, without any surface features larger than 0.1 mm, while "T" means macro-textured, with protruding and indented truncated cones in a regular square array. PTFE is poly(tetrafluoroethylene) and HFS is hydrophobic fumed silica.

| Surface identifier | Macro-texture ($>0.1$ mm) | Spray coating | Water advancing contact angle $\theta_a^*$ (°) | Water receding contact angle $\theta_r^*$ (°) |
|---|---|---|---|---|
| S-PTFE | No | PTFE | $165 \pm 3$ | $162 \pm 2$ |
| T-PTFE | Yes | PTFE | $166 \pm 1$ | $163 \pm 2$ |
| S-HFS | No | HFS | $163 \pm 4$ | $155 \pm 6$ |
| T-HFS | Yes | HFS | $167 \pm 2$ | $156 \pm 2$ |

We performed the droplet impact experiments (see Figure S2 for a sketch of the experimental setup) by dispensing a droplet from a height, $h$, with a syringe pump (Harvard Apparatus, Pump 33) and needle (Eppendorf Microloader 20L). In the case of pure water, the droplet radius before impact was $r_0 \approx 1.1$ mm. By adjusting $h$, we controlled the impact velocity, $u_0$, and with it the Weber number, $We = \rho u_0^2 r_0 / \gamma$, where $\rho$ and $\gamma$ are the density and the surface tension of the fluid, respectively. We recorded the droplet impact process with a high-speed camera (Photron, FastCam SA1.1) at a rate of 5,000 to 10,000 s$^{-1}$ using illumination from the back side (advanced illumination SL073). With MATLAB we computed $u_0$ from the recorded images.

**Results and Discussion**

First, we performed droplet impact experiments with pure, deionized water on all four samples: S-PTFE, T-PTFE, S-HFS, and T-HFS. For each experiment, we recorded the droplet-substrate contact time, $t_C$, and present the results in Figure 2(a). Figure 2(b) shows a typical image sequence of a droplet impacting on the S-PTFE surface. The impacting droplet undergoes spreading, retraction and rebound after a contact time of $t_C / \tau \approx 2.7$, where $\tau = \sqrt{\rho \cdot r_0^3 / \gamma}$ is the inertial-capillary time. Depending on $We$, $t_C / \tau$ ranges between 2.5 and 3.5 on the smooth surfaces. Figure 2(c) and (d) show image sequences of droplets impacting on the T-PTFE surface at a low and a high Weber number, respectively. For $We \leq 7$, we observe conventional rebound, while for $7 < We < 20$, we observe a rebound with reduced contact time $t_C / \tau \approx 1$ in a pancake shape. The pancake bouncing phenomenon was first described by Liu and co-workers.[9] They explained that when parts of the impacting droplet penetrate into specially designed surfaces, capillary energy is stored and subsequently rectified back in kinetic energy of the droplet. It is important that the stored energy is sufficient to lift the droplet. This first criterion sets the lower limit in Weber number for pancake bouncing, which we observe on our surface architecture to be at $We \approx 7$. The second necessary condition for pancake bouncing is that the time for the liquid to penetrate and empty the macro-texture, $t_\uparrow$, matches the time the droplet takes to spread out to its maximum diameter, $t_{max}$, which means that the ratio of these two times $k = t_\uparrow / t_{max}$ should be close to unity. On a surface with tapered posts, $k \sim w / \left(r_0 \sqrt{-\beta \cos \theta^*_{a/r}}\right)$ (Ref. 9). For our surface, $w / \left(r_0 \sqrt{-\beta \cos \theta^*_{a/r}}\right) \approx 0.5$, where $w = 0.34$ mm is the closest center to center distance between two protruding macro features in the square lattice, $\beta = 0.4$ is the difference between the base diameter and the top diameter divided by the cone height and $\theta^*_{a/r} = 160°$ is the effective contact angle. It is important to note that this analysis applies only when viscous effects can be neglected. Furthermore, due to its definition based on the time for the liquid to penetrate and empty the macro-texture, $t_\uparrow$, $k$ is only applicable to macro-textured surfaces. For $We > 20$ we observe incomplete rebound for both the S-PTFE and the T-PTFE surfaces. We conclude that the impalement resistance of the PTFE coating is not sufficient for $We > 20$ (see Figure S3 for side-view image sequences at $We > 20$ on the S-PTFE and the T-PTFE surface). In contrast, for $We > 20$ the droplets still fully rebound from the S-HFS surfaces, as the HFS coating is optimized for best impalement resistance. Interestingly, impacting droplets on T-HFS surfaces do not completely rebound, but liquid remains stuck in the macro-texture of the surface. We hypothesize that this is due to pinning at the sharp edges of the indentations of the macro-textured surfaces. The indentations were included in the surface design to allow for a mechanically stable architecture with short protrusions while still providing a large available surface area to store surface energy during droplet impact. Here we found that for $We > 20$ these indentations simultaneously increase the risk of pinning during the droplet retraction stage, which slows down the receding liquid within the macro-texture and can result in droplet breakup in the macro-texture (see Figure S3 for side-view image sequences at $We > 20$ on the S-HFS and the T-HFS surface). To facilitate dewetting and limit pinning events, it is important to reduce the presence of sharp edges, thus preventing liquid from remaining in the texture. This way the addition of macro-texture may be beneficial for an even broader range of values of $We$.

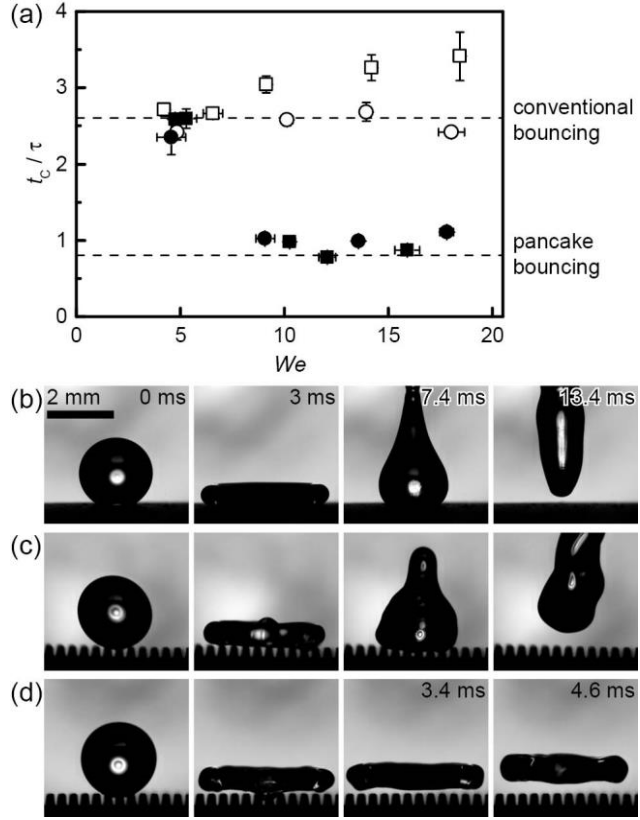

**Figure 2. Contact times of water droplets.** (a) Contact time $t_C$ divided by the inertial-capillary time $\tau = \sqrt{\rho \cdot r_0^3 / \gamma}$ vs Weber number $We$ for a droplet impacting the macroscopically smooth (PTFE coated: open squares; HFS coated: open circles) and the macro-textured surface (PTFE coated: filled squares; HFS coated: filled circles). Dashed horizontal lines show $t_C / \tau = 0.8$ and $t_C / \tau = 2.6$. $n \geq 4$ experiments per data point. Error bars show the standard deviation. (b) Conventional rebound on the S-PTFE surface at $We = 9.1$ (Video S1). (c) Conventional rebound on the T-PTFE surface at $We = 5.4$ (Video S2). (d) Pancake rebound on the T-PTFE surface at $We = 11.4$ (Video S3). All images have the same scale bar and are synchronized in time with (b) if not otherwise stated. For $We > 20$ we observe partial rebound on the macro-textured surfaces and the S-PTFE surface (see Figure S3).

For $We \approx 10$, where we observed highly repeatable pancake bouncing with water droplets not leaving any liquid behind on the surface, we proceeded to study how increasing the viscosity of the droplet affects the rebound process. To this end, we prepared eleven different water-glycerol mixtures ranging from pure water (low viscosity) to pure glycerol (high viscosity) in 10 wt.% increments. See Table S1 for how the density, surface tension and viscosity change as a function of the glycerol concentration in the mixture. In Figure 3(a) we plot $t_C / \tau$ vs the Ohnesorge number $Oh = \eta / \sqrt{\rho r_0 \gamma}$, which summarizes all results obtained with viscous droplets impacting on the T-PTFE surface and the S-PTFE surface. Here $r_0$ is the initial droplet radius and $\eta$ is the droplet dynamic viscosity. We performed the experiments on viscous pancake bouncing using the PTFE-coated samples as the PTFE is able to repel glycerol, in contrast to the HFS-coating, which is optimized for water repellency but cannot repel glycerol. We found that for water-glycerol mixtures ranging from 0 to 40 wt.% glycerol, impacting viscous droplets are repelled in a pancake shape by the T-PTFE surface architecture for $We \approx 10$. These water-glycerol mixtures correspond to a range of $1\,\text{mPa}\cdot\text{s} \leq \eta \leq 3.7\,\text{mPa}\cdot\text{s}$ and $0.003 \leq Oh \leq 0.013$. For comparison, supercooled water at $-17\,°\text{C}$ has $\eta = 3.7\,\text{mPa}\cdot\text{s}$ (Ref. 15). Figure 3(b) shows an image sequence of a droplet consisting of 60 wt.% water and 40 wt.% glycerol impacting on the T-PTFE surface and rebounding in a pancake shape. At a concentration of 50 wt.% glycerol in the aqueous mixture, which corresponds to $\eta_{\text{trans}} = 6\,\text{mPa}\cdot\text{s}$ and $Oh_{\text{trans}} = 0.021$, we observed a transition from pancake bouncing to conventional (stretch-recoil) rebound of the impacting droplet. For higher concentrations of glycerol ranging from 60 wt.% to 80 wt.% glycerol in the aqueous mixtures, which corresponds here to a range of $0.04 \leq Oh \leq 0.21$, all droplets undergo conventional rebound from the T-PTFE surface. Figure 3(c) shows the conventional rebound of a droplet consisting of an aqueous solution with 80 wt.% glycerol. In contrast to our observations on the T-PTFE surface, impacting droplets for water-glycerol mixtures ranging from 0 to 70 wt.% glycerol in the aqueous solution were undergoing conventional rebound on the S-PTFE surface. Interestingly, at high viscosities (80 wt.% glycerol in the aqueous solution), droplets were no longer bouncing on the S-PTFE surface, but remained immobile on it after impact as shown in Figure 3(d). Thus, the T-PTFE architecture provides both drastically reduced contact times for viscous droplets up to $Oh \leq 0.013$ and improved repellency of impacting droplets at $Oh = 0.21$ compared to the S-PTFE reference surface. In Figure 3 we investigated droplet impact on the T-PTFE surface as a function of liquid viscosity keeping a constant $We \approx 10$. To further study the effect of $We$ on the pancake bouncing regime, we performed a series of droplet impact experiments this time varying $We$ keeping the viscosity of the liquid constant by using an aqueous solution with 40 wt.% glycerol, which is the most viscous liquid that showed pancake bouncing at $We \approx 10$. We summarize our findings in Figure S4. We found that also at this level of viscosity the lower transition between conventional bouncing and pancake bouncing is at $We \approx 7$. For $We$ ranging between 8 and 12 we observed robust pancake bouncing. For $We = 14$ and above we observed partial rebound with small amounts of liquid remaining in the macro-texture.

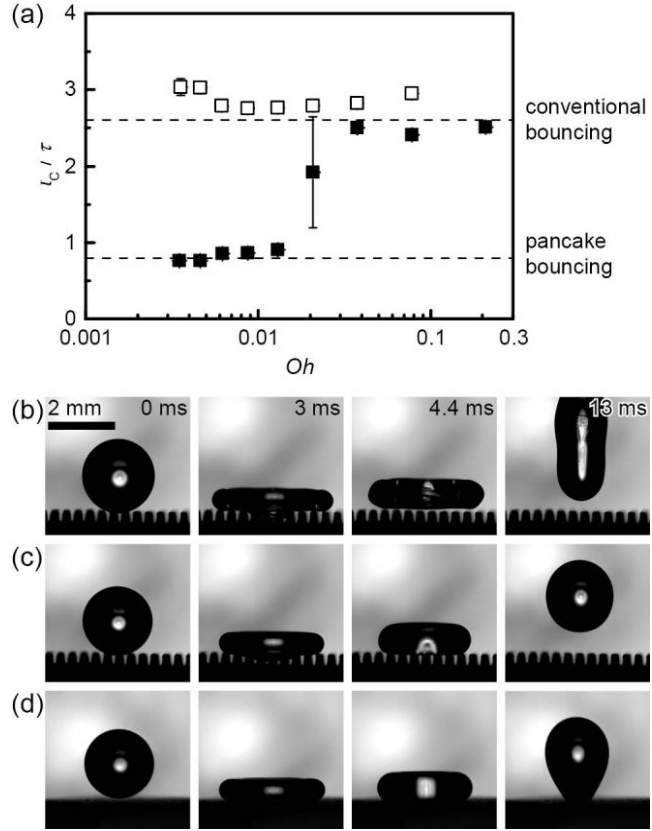

**Figure 3. Contact times of viscous droplets.** (a) Contact time $t_C$ divided by the inertial-capillary time $\tau = \sqrt{\rho \cdot r_0^3 / \gamma}$ vs Ohnesorge number $Oh$, for droplets impacting on the S-PTFE (open squares) and the T-PTFE (filled squares) surfaces. Horizontal lines are at $t_C / \tau = 0.8$ and $t_C / \tau = 2.6$. (b) Pancake rebound on the T-PTFE surface at $Oh = 0.013$ (Video S4). (c) Conventional rebound on the T-PTFE surface at $Oh = 0.21$ (Video S5). (d) No rebound on the S-PTFE surface at $Oh = 0.21$ (Video S6). $We \approx 10$ for all experiments. $n \geq 4$ experiments per data point. Error bars show the standard deviation.

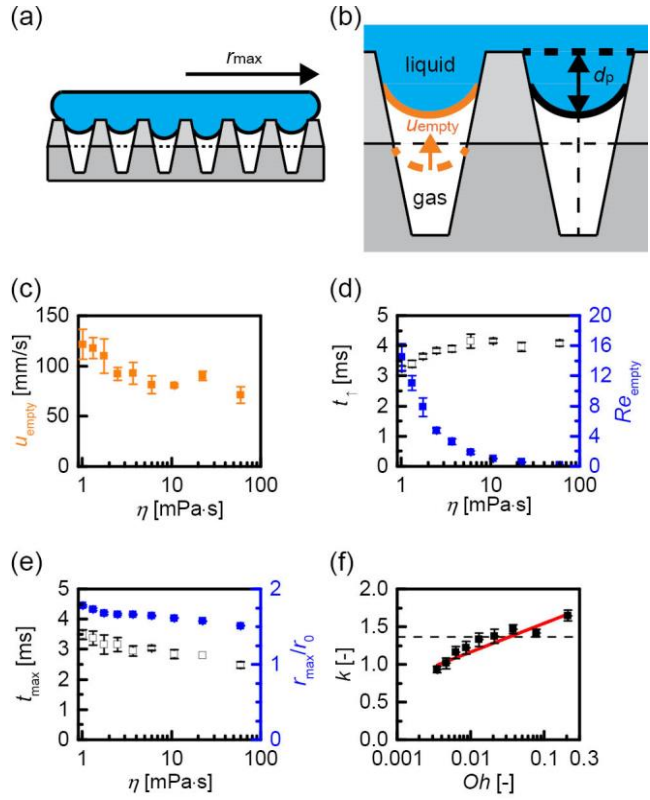

**Figure 4. Fluid flow analysis of viscous droplets impacting on the macro-textured surface.** (a) Side-view schematic of a droplet impacting on the macro-textured surface introducing the radius at maximum spread $r_{max}$. (b) Zoomed in section-view schematic showing the fluid emptying the texture, introducing the average fluid velocity during the capillary emptying process $u_{empty}$ and the average fluid penetration depth $d_p$. (c) Experimentally measured $u_{empty}$ vs fluid viscosity $\eta$. (d) Time $t_\uparrow$ for the liquid to penetrate and empty the texture (open black squares) and effective Reynolds number $Re_{empty}$ the fluid experiences when emptying the texture (filled blue squares) vs $\eta$. (e) Time $t_{max}$ (open black squares) to reach the maximum spreading radius, and $r_{max}$ divided by the initial droplet radius $r_0$ (filled blue squares) vs $\eta$. (f) Ratio $k = t_\uparrow / t_{max}$ vs $Oh$. The dashed horizontal line introduces the level of $k_{trans}$ at which we observe the transition from pancake bouncing to conventional bouncing. The red line shows a fit to the experimental data with $k = 0.381 \cdot \log(Oh) + 1.93$. $We \approx 10$ for all experiments. $n \geq 4$ experiments per data point. Error bars show the standard deviation.

To understand why the macro-textured surface architecture can repel viscous droplets in a pancake shape up to this specific $\eta_{\text{trans}}$ and $Oh_{\text{trans}}$, we analyzed the fluid flow of the impacting droplet above and within the texture. Figure 4(a) shows a schematic of the droplet on the surface texture during maximum spread with the radius $r_{\text{max}}$. Figure 4(b) shows schematically the capillary emptying process, introducing the average fluid velocity when the droplet empties the texture $u_{\text{empty}}$ and the effective average penetration depth of the fluid into the texture $d_{\text{P}}$. We assessed $u_{\text{empty}}$ experimentally for all droplet impact events by analyzing the side-view high-speed images. We first measured the volume of the portion of the droplet that did not penetrate the texture but remained above the texture during the droplet impact event *vs* time. Therefore, we binarized the recorded side-view images, detected the top of the surface texture and determined the centroid of the droplet from the side-view perspective. On this basis and by assuming rotational symmetry of the droplet, we computed the volume of the portion of the droplet remaining above the texture. Subtracting the droplet volume above the texture from the initial droplet volume (measured before impact) yields the volume of the remaining portion of the droplet that penetrated the texture. Simultaneously, we measured the contact area of the droplet at the moment of maximum penetration from the side-view high-speed images. Based on these we computed $d_{\text{P}}$, which we define as the maximum of the penetrated droplet volume divided by the contact area at this moment, finding $d_{\text{P}}$ to be in the range of $\approx 0.2$ mm for all values of $\eta$. We also measured the time the fluid takes from the moment of maximum penetration to empty the texture $t_{\text{empty}}$. Based on these we computed $u_{empty}$ as $u_{empty} = d_P / t_{empty}$. In Figure 4(c) we plot $u_{empty}$ vs $\eta$, which shows that the more viscous the droplet, the slower it empties the texture. In Figure 4(d) we plot $t_\uparrow$ vs $\eta$, finding that $t_\uparrow$ rises with $\eta$. To explain this trend, we show the effective Reynolds number during texture emptying $Re_{empty} = \rho u_{empty} r_{open} / \eta$ in the same graph. We see that when alternating the fluid composition from pure water to a mixture of water with 50 wt.% glycerol, which is the amount of glycerol where the transition between pancake bouncing and conventional rebound in Figure 3(a) occurs, the viscosity increases by 600 %, while the fluid density $\rho$ changes by less than 15 %. Consequently, $Re_{empty}$ reduces with increasing glycerol concentration from $Re_{empty} = 14$ for pure water, to $Re_{empty} = 1.8$ for the 50 wt.% glycerol aqueous solution, and down to $Re_{empty} = 0.2$ for the 80 wt.% glycerol aqueous solution. This supports the conclusion that viscous effects become important at 50 wt.% glycerol concentration as here $Re_{empty}$ approaches unity and explains why we observe viscous slow down and the significant increase in $t_\uparrow$. For larger glycerol concentrations, the fluid motion, as the droplet empties the surface texture, is in a capillarity-viscosity regime, where the droplet retraction rate is inversely proportional to $\eta$ (Ref. 17). In this context in is evident that increasing $\eta$ results in increasing $t_\uparrow$, as observed in Figure 4(d). In Figure 4(e) we plot $t_{max}$ vs $\eta$, finding that $t_{max}$ decreases with increasing $\eta$. To explain this behavior, we plot the maximum spreading radius divided by the initial droplet radius $r_{max} / r_0$ in the same graph. We find that $r_{max} / r_0$ decreases with $\eta$ for a similar impact speed $u_0$, which agrees with the theory.[18,19] The time $t_{max}$ that a droplet requires during an inertia-capillarity dominated impact to reach $r_{max}$ scales with $t_{max} \sim \sqrt{\rho r_{max}^3 / \gamma}$ (Ref. 20). Combining the two previous statements yields that as $\eta$ increases, $t_{max}$ has to decrease, as observed in Figure 4(e). In Figure 4(f) we summarize all previous findings into one graph showing $k = t_\uparrow / t_{max}$ vs $Oh$. The two trends, which we previously described separately, namely the increase of $t_\uparrow$ with increasing $\eta$ and the decrease of $t_{max}$ with increasing $\eta$ now combine into a single effect. As $\eta$ and with it $Oh$ rises, $k$ rises. We fit a curve to the data which yields $k = 0.381 \cdot \log(Oh) + 1.93$. In this plot we distinguish two regions. The first region is the pancake bouncing region including all glycerol concentrations from 0 wt.% up to 40 wt.% glycerol in the aqueous solution, which translates here into $0.003 \leq Oh \leq 0.013$. For pure water with $Oh = 0.003$ we measure $k \approx 0.93$, while for an aqueous solution with 40 wt.% glycerol with $Oh = 0.013$ we find $k \approx 1.33$. For these mixtures we observe that the droplet empties the texture at a time when it is at its maximum spread, so that here $t_\uparrow \approx t_{max}$ (see Figure 3(b), VideoS4). Previous work has shown that in this range of $0.003 \leq Oh \leq 0.013$ both impact and rebound dynamics on smooth surfaces are governed by a competition of inertia and capillarity, while viscous effects are of minor importance.[17] A further increase of the glycerol concentration up to 50 wt.% and with it an increase of $Oh$ to 0.021 results in a transition from pancake bouncing to conventional rebound, as observed in Figure 3(a). The mismatch between $t_\uparrow$ and $t_{max}$ has increased to an extend that their ratio $k$ reached the critical value of $k_{trans} \approx 1.37$ no longer allowing for pancake bouncing. By further increasing the glycerol concentration and with it $Oh$ to $0.037 \leq Oh \leq 0.21$ we reach the second region, where $k$ rises above $k_{trans}$ up to 1.65 for the case of 80 wt.% glycerol. Based on our previous Reynolds number analysis in Figure 4(d) we expect that in addition to inertia and capillarity, in this second region viscosity affects the receding dynamics, slowing down the drainage flow within the surface texture.[17] The combined effect of increasing $t_\uparrow$ and decreasing $t_{max}$ with increasing $\eta$ for a fixed surface texture results in an increasing mismatch between $t_\uparrow$ and $t_{max}$, and consequently the transition from pancake bouncing to conventional bouncing. In addition to this mismatch, increasing the droplet viscosity to these critical levels also increases the energy consumed by viscous dissipation, which reduces the available energy for rebound. In order to reduce the mismatch between $t_\uparrow$ and $t_{max}$, one has to limit the viscous slow down inside the texture, which can be achieved by increasing the macro-feature size or their distance, so that $Re_{empty}$ is increased and viscous effects leading to the slow down are reduced. Simultaneously, it has to be considered, that there is a practical upper limit to increasing the feature size as the features have to still be markedly smaller than the impacting droplets for pancake bouncing to occur irrespective of the precise impact position on the surface texture. This can be a design criterion for the development of next generation macro-textured surfaces with enhanced viscous droplet repellency.

**Conclusions**

We have shown that 3D printing can be used to generate, in a facile manner, macro-textured surface architectures with alternating, protruding and indented truncated cones of the size of the order of $0.1\,\text{mm}$, which are mechanically robust due to their short size while simultaneously providing a large surface area. Combined with soft lithography and spray coating we fabricated superhydrophobic surfaces that can not only repel water droplets with reduced contact time in a pancake shape, but also significantly more viscous droplets with a fluid viscosity of up to $3.7\,\text{mPa}\cdot\text{s}$, which is more than three times the value of pure water at room temperature. We determined the viscous limitation of pancake bouncing on our surface and explained the transition from pancake bouncing to conventional bouncing by a viscosity dependent mismatch between the time for the droplet to spread to its maximum diameter and the time for the liquid to fill and empty the surface topography. We propose that larger features can help to reduce viscous slow down during capillary emptying of the texture. Furthermore, we showed that such macro-textured surface architectures can shed highly viscous droplets, which cannot rebound from lotus leaf-like surfaces.

**Acknowledgements**

We acknowledge I. Karlin, S. S. Chikatamarla and A. M. Mazloomi for suggesting the surface design, as well as fruitful discussions. We thank U. Drechsler, S. Krödel, T. Vasileiou and A. Milionis for advice on surface fabrication; and J. Vidic and P. Feusi for assistance in experimental setup construction. Partial support of the European Research Council under Advanced Grant 669908 (INTICE) and the Swiss National Science Foundation under Grant 162565 is acknowledged.

**Author contributions**

D.P. and T.M.S. designed research; G.G. and O.B.M.K. performed research; G.G. and T.M.S. analyzed data; and G.G., T.M.S. and D.P. wrote the paper.

**Competing financial interests**

The authors declare no competing financial interests.

**Supporting Information**

The Supporting Information is available free of charge online on the ACS Publications website. It contains: Figure S1. Characteristics of the T-HFS surface; Figure S2. Schematic of the experimental setup; Figure S3: Water droplet impact at high Weber numbers; Figure S4: Viscous droplet impact on the macro-textured, PTFE-coated surface using an aqueous solution with 40 wt.% glycerol; Table S1. Physical properties of the water glycerol mixtures (pdf). It also contains six Supporting Information Videos: Video S1. Conventional rebound of a water droplet impacting on the S-PTFE surface at $We = 9.1$; Video S2. Conventional rebound of a water droplet impacting on the T-PTFE surface at $We = 5.4$; Video S3. Pancake rebound of a water droplet impacting on the T-PTFE surface at $We = 11.4$; Video S4. Pancake rebound of a droplet with 60 wt.% water and 40 wt.% glycerol impacting on the T-PTFE surface at $Oh = 0.013$; Video S5. Conventional rebound of a droplet with 20 wt.% water and 80 wt.% glycerol impacting on the T-PTFE surface at $Oh = 0.21$; Video S6. No rebound of a droplet with 20 wt.% water and 80 wt.% glycerol impacting on the S-PTFE surface at $Oh = 0.21$.

**Graphical TOC**

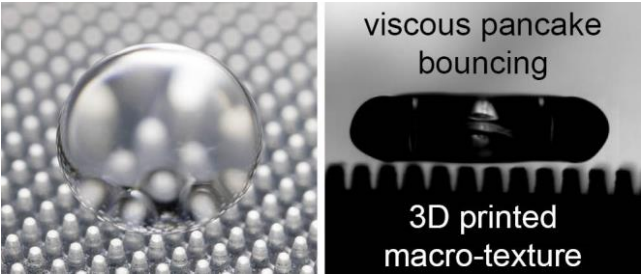